\documentclass{article}

\usepackage{arxiv}

\usepackage[utf8]{inputenc} % allow utf-8 input
\usepackage[T1]{fontenc}    % use 8-bit T1 fonts
\usepackage{hyperref}       % hyperlinks
\usepackage{url}            % simple URL typesetting
\usepackage{booktabs}       % professional-quality tables
\usepackage{amsfonts}       % blackboard math symbols
\usepackage{nicefrac}       % compact symbols for 1/2, etc.
\usepackage{microtype}      % microtypography
\usepackage{graphicx}
\usepackage{natbib}
\usepackage{doi}
\usepackage{amsmath}
\usepackage{amsthm}

\newtheorem{Proposition}{Proposition}

\title{Digital leisure and the gig economy:\\ a two-sector model of growth}

\author{ \href{https://orcid.org/0000-0002-7903-964X}{\includegraphics[scale=0.06]{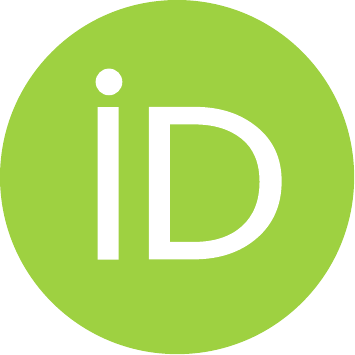}\hspace{1mm}Francesco~Angelini}%\thanks{Use footnote for providing further
\\
	Department of Statistical Sciences Science ``Paolo Fortunati''\\
	University of Bologna\\
	47921 Rimini, IT \\
	\texttt{francesco.angelini7@unibo.it} \\
	\And
	\href{https://orcid.org/0000-0001-7205-6319}{\includegraphics[scale=0.06]{orcid.pdf}\hspace{1mm}Luca V.~Ballestra} \\
        Department of Statistical Sciences Science ``Paolo Fortunati''\\
	University of Bologna\\
	47921 Rimini, IT \\
	\texttt{luca.ballestra@unibo.it} \\
 	\And
	\href{https://orcid.org/0000-0002-5814-5902}{\includegraphics[scale=0.06]{orcid.pdf}\hspace{1mm}Massimiliano~Castellani} \\
	Department of Statistical Sciences Science ``Paolo Fortunati''\\
	University of Bologna\\
	47921 Rimini, IT \\
	\texttt{m.castellani@unibo.it} \\
}

\renewcommand{\shorttitle}{Digital leisure and the gig economy}

\hypersetup{
pdftitle={Digital leisure and the gig economy: a two-sector model of growth},
pdfsubject={econ.TH},
pdfauthor={Francesco~Angelini, Luca V.~Ballestra, Massimiliano~Castellani},
pdfkeywords={Digital leisure, Data, Endogenous growth, Gig Economy},
}

\begin{document}
\maketitle

\begin{abstract}
The process of market digitization at the world level and the increasing and extended usage of digital devices reshaped the way consumers employ their leisure time, with the emergence of what can be called digital leisure. This new type of leisure produces data that firms can use, with no explicit cost paid by consumers. At the same time, the global digitalization  process has allowed workers to allocate part of (or their whole) working time to the Gig Economy sector, which strongly relies on data as a production factor. In this paper, we develp a two-sector growth model to study how the above mechanism can shape the dynamics of growth, also assessing how shocks in either the traditional or the Gig Economy sector can modify the equilibrium of the overall economy. We find that shocks in the TFP can crowd out working time from a sector to the other, while shocks on the elasticity of production to data determines a change in the time allocated to digital leisure.
\end{abstract}

% keywords can be removed
\keywords{Digital leisure \and Data \and Endogenous growth \and Gig Economy}

\section{Introduction}\label{sectionintro}

Recently, the world economy witnessed a process of digitalization of the market, fostered by the emergence of digital platforms, the diffusion of digital devices making it possible to use digital services very easily, and the highest availability of data that allowed firms to serve the demand better and to create new products. This process affected people's consumption habits, work schedules, as well as the way they manage their leisure time, and introduced new job types \citep{bearson2021}.\\
The emergence and diffusion of the gig economy (hereafter GE) have strongly been fostered by the digital revolution, and one may argue that GE is itself part of this revolution, affecting people's labor and leisure choices. Among the different tasks and activities typical of the GE, we can identify a series of “additional jobs”, such as subletting a room or an apartment, using one’s private car to drive persons around or bring food and goods to the customers of a firm, or providing digital services from remote \citep{wood2019}. We say “additional jobs” because they allow persons to supplement the income of their principal job as if they were self-employed workers \citep{abraham2021}. Thus, GE activities may erode the leisure time that citizens have at their disposal, and can be the main source of income for some categories of persons \citep{wu2019}.\footnote{See \citet{poon2019} for an overview of growth trends of GE workers and \citet{deranty2022} for a critical review of platform work in social sciences.} 

The digitalization process has also impacted the way leisure time is managed. Important technological innovations have already been proven to influence the way people organize their free time \citep{rachel2021}. Key examples of this are, in the past, the introduction of TV or the phonograph and, more recently, the advent of the computer and the Internet \citep{bryce2001,aguiar2007} and the smartphones, whose usage is typical of several people's leisure time and also shapes the way they enjoy it \citep{lepp2013,lepp2014,ciochetto2015}. However, the changes brought by the diffusion of digital technologies have a different impact on people's choices with respect to previous technological innovations \citep{lopezsintas2017}. In particular, \citet{allcott} report evidence that social media usage is habit forming and their users possibly present self-control problems.\footnote{The use of social networking sites has been found to have a negative impact on users' well-being, reducing their social trust, even though it also has a positive impact on well-being, since it increases the probability of face-to-face interactions \citep{sabatini2017}.}\\
Consumers may believe that the leisure time they spend using smartphones or other devices that allow access to the Internet has no cost (neglecting the cost of connecting to the network), but, actually, they are paying for their enjoyment since they share their data. Indeed, data can be seen as a by-product of this type of consumption, being a central production factor for both digital and non-digital firms in the last years \citep{jones2020}. So, data is not directly paid by firms to the consumers, but indirectly through the supplied services \citep{acemoglu2019}, in exchange for the possibility for firms to better target their audiences \citep{farboodi2019growth,bergemann2022}, to improve the quality of their products \citep{schaefer2019}, and to obtain information from uncertain states of the world \citep{ichihashi2021}.\footnote{The continuous production of big data made by the consumers suggests that data might not be a capital stock but another good resulting from consumers' labour, which is not paid or at least is not paid in an ordinary way \citep{arrieta2018}. It has to be highlighted that data can also be created during the job time, which especially happens for the GE additional jobs we mentioned previously \citep{attoh2019}.}

Focusing on the impact of GE on labor and leisure and the consequent change in the way people make their choice of consumption, in this paper, we investigate how the digitization of the market and the emergence of GE influence economic growth in the long run. To this aim, we develop an economic growth model with two production sectors, namely, a physical and a digital sector, taking also into account time allocation choices and the role of data as a by-product of the leisure activities of consumers. Such a framework also allows us to study how digital innovation, e.g., shocks on the production function or an increased ability in making agents produce data, can modify the economic equilibrium. Furthermore, we can also investigate how shocks impact the allocation of time among traditional jobs, GE jobs, and leisure.

We show that a shock on the total factor productivity in a sector of the economy increases the working time allocated in that sector and decreases the working time spent in the other sector, whereas the time spent on digital leisure does not substantially vary. Furthermore, we also highlight how a technological change in the elasticity of production to data has a positive effect on both leisure time and working time in the digital sector. This analysis provides a novel result: the time allocated to digital leisure and to the job in the GE sector crowds out the working time spent in the physical sector as the importance of data in the GE sector production increases.

Our paper contributes to the literature that investigates the relationship between data and economic growth and, in particular, to the emerging strand that considers data as a production factor. In particular, \citet{cong2021} developed an endogenous growth model where data is consumer-generated and stimulates innovation and knowledge accumulation. The authors take also into account the concerns that consumers may have about their privacy, showing how they may influence their data-sharing choices. In addition, \citet{yang2021} developed a two-sector growth model where each sector has a different level of data deepening and is characterized by a different amount of skilled workers (i.e., workers who can use data), finding that an increase in the stock of data increases the number of skilled workers in the data-intensive sector and reduces the difference between the wages of skilled workers and the wages of unskilled workers. \citet{jones2020} studied how data and ideas enter the production function, together with labor, and how they interact with consumption in an economy characterized by digital firms, where positive data externalities among firms are possible. \citet{farboodi2019growth} built a dynamic model where data is regarded as information that helps firms in their prediction and decision tasks and studied if the data-accumulation mechanism differs from the capital accumulation mechanism, finding that a difference exists only in the short term.\\ 
Among this emerging strand of literature, the paper that is most related to our analysis is that by \citet{rachel2021}, who studies leisure-enhancing technologies (i.e., technologies that increase the enjoyment resulting from leisure time, generally free of charge, obtained in exchange of time and attention), their impact on growth, and their coexistence with traditional innovation. The paper finds that the number of hours worked decreases if leisure-enhancing innovations are available, and also suggests that GDP could get more distant from actual welfare if more leisure-enhancing technologies were available, which is a possible explanation for the low growth revealed by the empirical data. What our paper considers, which is not accounted for by \citeauthor{rachel2021}'s model, is the existence of GE and the usage of data by firms in this sector; by considering the mechanism through which data affects production in the GE sector we can study how the choices of allocating times among working in the physical sector, working in the digital sector, and leisure impact economic growth.\\
Furthermore, and more in general, what is still missing in the strand of papers considering the interaction between growth and data is a focus on digital leisure, which can indeed be an activity that generates a flow of data relevant to firms. This kind of digital activity has been investigated in other contexts, mostly regarding how consumers' choices impact social capital accumulation when facing the choice between digital and face-to-face interaction \citep{sodini2012,sodini2014}. An exception is the two-sector growth model by \citep{guerrini}, who study how quality leisure time, which can be affected by choice of digital consumption, can impact growth in a two-sector setting.

Our paper also adds to the existing literature on leisure and growth. Taking an endogenous growth and a real business model perspective, \citet{benhabib1991}, \citet{greenwood1991}, \citet{einarsson1997}, and \citet{perli1998} investigated the use of non-market time with different approaches, exploiting a household production technology based on non-market work and household capital. In particular, this literature is capable to disaggregate the work activities and the capital employed in the household sectors and in the market. Along a similar line, \cite{candela2019} generalized the endogenous growth framework of \cite{lad1999} by assuming that leisure time includes both ``pure leisure'' and time spent on the consumption of marketable leisure services. In particular, they disaggregated leisure time into non-marketable and marketable time, where the time spent jointly with marketable services governs the dynamic of an additional production sector such as the service sector. Another strand of literature concerning the ``disaggregation'' of leisure shows that different ``types'' of leisure may have a different impact on economic activities when a share of leisure time is spent jointly with marketable services \citep{aguiar2007,ngai2008,vandenbroucke2009,boppart2018}.
Among this literature, \citet{boikos2022} is the paper that is most related to our work. The authors studied if and how leisure can impact productivity, framing their analysis in a horizontal R\&D-based growth model, assuming that leisure externalities are at work and impact the productivity of researchers. They found that these externalities positively affect the production of new ideas and hence stimulate R\&D if the stock of previous ideas does not generate too strong spillover effects on this production. In line with \citet{boikos2022}, our model both includes a ``productive leisure'' and takes into account how leisure choice is influenced by the emergence of the GE and how this has an impact on growth. In addition, our approach also allows us to study how productivity shocks can shape time allocation choices by consumers.

\section{The growth model}\label{sectionmodel}
In our setting, individuals consume physical goods and use digital (internet) services, both positively contributing to their utility. Individuals are assumed to be homogeneous, i.e. we can consider a representative agent who consumes an amount $c$ of physical goods (coming from both the physical sector and the GE sector) and enjoy digital leisure by spending an amount $d$ of time in digital services. Among the former, we have all goods sold by either a brick-and-mortar trader or a digital trader (such as the goods bought through smartphone apps), $d$ represents the actual time spent on the Internet using digital platforms, which is in and by itself a source of utility. In other words, digital leisure has a positive impact on the user's utility.\footnote{To simplify the model, we do not explicitly take into account the possibility of enjoying non-digital leisure in our framework. However, one can argue that, even when people do not spend their time directly using digital services in their free time, they still produce data, for example, related to resting time recorded through smartwatches or smartphones, or GPS position when not moving, etc. These data can also be produced during active non-digital leisure time, as in the case of apps recording the track during a hike or a run, or getting health information during similar activities.}\\
The utility function of the representative agent is assumed to be as follows:

\begin{eqnarray}\label{uagent}
u(c,d) = \frac{1}{1-\sigma} \left(c^{\sigma_1} d^{\sigma_2} \right)^{1-\sigma}, \,\,\,\,\, \sigma_1>0, \, \sigma_2>0, \,\,\,\,\, \sigma_1 + \sigma_2 = 1, \,\,\,\,\, \sigma>0, \, \sigma \neq 1.
\end{eqnarray}

This formulation allows for some degree of substitution between goods and digital services consumption.

Let $N(t)$ denote the total number of workers in the economy at time $t$. Following a common approach \citep{kelley1988}, we assume that $N(t)$ grows exponentially as in $N(t) = N_0 e^{nt}$, with $n>0$. 
Given that, the lifetime discounted utility of the representative agent is:
\begin{eqnarray}\label{totalutility}
U = \int_0^{+\infty} N(t) u(c(t),d(t)) e^{-\rho t} dt= N_0 \int_0^{+\infty} u(c(t),d(t)) e^{(n -\rho) t} dt.
\end{eqnarray}
where $\rho$ is the intertemporal discount rate and is assumed to be greater than $n$ to ensure that $U$ is finite.

Each consumer in our model has a total time (normalized to unity) he has to share between working time in the physical sector ($h_p$), working time in the GE sector ($h_d$), and using digital services as digital leisure time ($d$). These three values must be non-negative and sum to 1.

\begin{eqnarray}\label{constrainth}
h_p \geq 0, \,\,\,\,\,\,\, h_d \geq 0,\,\,\,\,\,\,\, d\geq 0, \,\,\,\,\,\,\, h_p + h_d + d = 1.
\end{eqnarray}

While the time worked in the physical and GE sector works as in traditional production function models, in our case we also consider the role of digital service usage, which produces a key factor for digital firms. In other words, the consumer who (freely) consumes digital services produces data that becomes an input for GE firms. Data work in fact as a revenue shifter, since firms can use them to improve and/or customize their products \citep{farboodi2019growth}, possibly being able to increase their prices and/or their sold quantity. Data can also be used to improve the advertising strategy of a firm potentially starting a virtuous circle, where ads influence in turn customers' choice towards the firm's products \citep{martins2019}. In our setting, we let the data produced through digital consumption impact only those firms more active in the digital sector, namely those who make use of digital platforms to sell their products and services, even though data could in principle be used also by firms in the physical sector.\footnote{Assuming that only digital-sector firms are able to properly use data as an input implicitly means that they have a better production function to transform data into a usable factor, a similar situation to those of firms using recycled products \citep{george2015}.} This translates to the following formulas for physical and GE production:\footnote{Notice that the functional form of \eqref{yp} is similar those of \eqref{yd} if we assume that the $h_dN$ has an exponent that tends towards zero (namely $b_2\rightarrow0$). This assumption can be explained by a lack of traditional-market firms' ability to use data produced during digital leisure as a production factor, or by a low efficiency in doing so leading to no impact on production whichever is the level of data produced by consumers.}

\begin{eqnarray}\label{yp}
Y_p = A_p (u_p K)^{a_1} (h_p N)^{a_2}, \,\,\,\, A_p>0, \,\,\,\, a_1>0, \,\,\,\, a_2>0,
\end{eqnarray}\begin{eqnarray}\label{yd}
Y_d = A_d (u_d K)^{b_1} (h_d N)^{b_2} (d N)^{b_3}, \,\,\,\, A_d>0, \,\,\,\, b_1>0, \,\,\,\, b_2>0, \,\,\,\, b_3>0
\end{eqnarray}

where $Y_p$ and $Y_d$ are the output of physical and GE firms, respectively, $K$ is the total capital employed in production, and $u_p$ and $u_d$ are the fractions of capital allocated to the physical and GE firms, respectively. Note that $A_p$ and $A_d$ denote the technology levels in the physical and GE sectors, capturing the part of production that is not due to the explicitly defined factors, and are assumed to be exogenous, contrary to what is done in other paper such as \citet{farboodi2019growth}.

Furthermore, production is assumed to have constant returns to scale, this implying

\begin{eqnarray}\label{cobbab}
a_1 + a_2 = 1, \,\,\,\,\,\,\, b_1 + b_2 + b_3=1.
\end{eqnarray}

Finally, let us define the total output as:
\begin{eqnarray}\label{Ytotal}
Y = Y_p + Y_d.
\end{eqnarray}
and the outputs and capital per capita:
\begin{eqnarray}\label{ycdpercapita}
y_p = \frac{Y_p}{N}, \,\,\,\,\,\, y_d = \frac{Y_d}{N}, \,\,\,\,\,\, y = \frac{Y}{N},
 \,\,\,\,\,\, k = \frac{K}{N}.
\end{eqnarray}

By using (\ref{cobbab}) and (\ref{ycdpercapita}), equations (\ref{yp}), (\ref{yd}) and (\ref{Ytotal}) are respectively rewritten as follows:

\begin{eqnarray}\label{ypsmall}
y_p = A_p (u_p k)^{a_1} h_p^{a_2},
\end{eqnarray}\begin{eqnarray}\label{ydsmall}
y_d = A_d (u_d k)^{b_1} h_d^{b_2} d^{b_3},
\end{eqnarray}\begin{eqnarray}\label{ytotalsmall}
y = y_p + y_d,
\end{eqnarray}

Moreover, by assuming that the capital depreciates at a constant rate $\delta>0$, the total capital per capita has to satisfy

\begin{eqnarray}\label{dynamicsk}
\dot{k} = A_p (u_p k)^{a_1} h_p^{a_2} + A_d (u_d k)^{b_1} h_d^{b_2} d^{b_3} - c - (\delta + n)k.
\end{eqnarray}

We choose the consumption $c$, the amount of times $d$, $h_p$ and $h_d$ and $u_p$ and $u_d$ that maximize the total utility (\ref{totalutility}).

To this aim, let us consider the Hamiltonian

\begin{eqnarray}\label{hamiltonian}
H(c,u_p,u_d,h_p,h_d,k,\lambda) = \frac{1}{1-\sigma} \left(c^{\sigma_1} d^{\sigma_2} \right)^{1-\sigma} + \lambda(A_p (u_p k)^{a_1} h_p^{a_2} + A_d (u_d k)^{b_1} h_d^{b_2} d^{b_3} \nonumber \\ [6pt]
- c - (\delta + n)k).
\end{eqnarray}

By keeping into account the constraints (\ref{constrainth}) and on $u_p$ and $u_d$, the first-order optimization conditions are as follows:

\begin{eqnarray}\label{Hc}
\frac{(1-\sigma)\sigma_1 u(c,1-h_p-h_d)}{c} - \lambda = 0,
\end{eqnarray}\begin{eqnarray}\label{Hup}
\frac{a_1 y_p}{u_p} - \frac{b_1 y_d}{u_d}= 0,
\end{eqnarray}

\begin{eqnarray}\label{Hhp}
\lambda \left(\frac{a_2 y_p}{h_p} - \frac{b_3 y_d}{1 - h_p - h_d}\right) - \frac{(1-\sigma)\sigma_2 u(c,1-h_p-h_d)}{1 - h_p - h_d} = 0,
\end{eqnarray}

\begin{eqnarray}\label{Hhd}
\lambda \left(\frac{b_2 y_d}{h_d} - \frac{b_3 y_d}{1 - h_p - h_d}\right) - \frac{(1-\sigma)\sigma_2 u(c,1-h_p-h_d)}{1 - h_p - h_d} = 0,
\end{eqnarray}

\begin{eqnarray}\label{Hlambda}
\dot{\lambda}
= - \lambda \left(
\frac{a_1 y_p + a_2 y_d}{k} -(\delta + \rho)
\right),
\end{eqnarray}

subject to the constraints (\ref{constrainth}), on $u_p$ and $u_d$, (\ref{ypsmall}), (\ref{ydsmall}),
 (\ref{ytotalsmall}) and (\ref{dynamicsk}).

We look for a steady state solution of system of equations (\ref{Hc})-(\ref{Hlambda}) with the constraints (\ref{constrainth}), on $u_p$ and $u_d$, (\ref{ypsmall}), (\ref{ydsmall}), (\ref{ytotalsmall}),
 (\ref{dynamicsk}), i.e. a constant solution $y_p = \overline{y}_p$, $y_d = \overline{y}_d$,
 $y = \overline{y}$,
 $h_p = \overline{h}_p$, $h_d = \overline{h}_d$, $d = \overline{d}$, $u_p = \overline{u}_p$,
 $u_d = \overline{u}_d$,
 $k = \overline{k}$, $c = \overline{c}$, $u = \overline{u}$, $\lambda = \overline{\lambda}$. Moreover, the solution we are looking for should both maximize the utility (\ref{totalutility}) and be economically meaningful, and thus we also require:\footnote{Let us observe that, given the growth rate of $N(t)$ and (\ref{ycdpercapita}), the steady state solution is such that the corresponding physical output $\overline{Y}_p$, digital output $\overline{Y}_d$, total output $\overline{Y}$ and total capital $\overline{K}$ grow exponentially, at a rate equal to $n$ (actually, the steady state solution is a balanced growth path).}

 \begin{eqnarray}\label{additionalrequirements}
 \overline{y}_p \geq 0, \,\,\,\,\,\,\, \overline{y}_d \geq 0, \,\,\,\,\,\,\, \overline{y} > 0,\,\,\,\,\,\,\,
 \overline{k} \geq 0, \,\,\,\,\,\,\, \overline{c} \geq 0, \,\,\,\,\,\,\, \overline{u}> 0.
 \end{eqnarray}

We have the following result.

\begin{Proposition}\label{propgrowthrate}
Let us define
\begin{eqnarray}\label{M1}
 M_1 = \left( \frac{A_p a_1^{a_1} a_2^{a_2}}{(\delta+\rho)^{a_1}} \right)^{\frac{1}{a_2}},
 \,\,\,\,\,\,\,\,
M_2 = \left( M_1^{b_2+b_3} \frac{(\delta+\rho)^{b_1}}{A_d b_1^{b_1} b_2^{b_2}}
\right)^{\frac{1}{b_3}},\,\,\,\,\,\,\,\,M_3 = 1-a_2\frac{\delta + n}{\delta + \rho} - M_2+b_3.
\end{eqnarray}
The system of equations (\ref{Hc})-(\ref{Hlambda}) with the constraints (\ref{constrainth}), on $u_p$ and $u_d$, (\ref{ypsmall}), (\ref{ydsmall}),
(\ref{dynamicsk}) has an economically meaningful steady state solution (satisfying (\ref{additionalrequirements}))
if and only if
\begin{eqnarray}\label{assumptionsM}
M_2 - b_3 >0, \,\,\,\,\,\,\,\,\,\,\, M_3 \leq 0.
\end{eqnarray}
Moreover, such a steady-state solution is unique and is given by
\begin{eqnarray}\label{ypbarydbar}
\overline{y}_p = -\frac{M_1 M_3}{\Delta}, \,\,\,\,\,\,\,\,\,\,\,
\overline{y}_d = \frac{M_1 P}{\Delta}, \,\,\,\,\,\,\,\,\,\,\,
\overline{y} = \frac{M_1(P-M_3)}{\Delta},
\end{eqnarray}
\begin{eqnarray}\label{hpbar}
\overline{h}_p = \frac{a_2 \overline{y}_p}{a_2 \overline{y}_p + (b_2+ M_2) \overline{y}_d}, \,\,\,\,\,\,\,\,\,
\overline{h}_d = \frac{b_2 \overline{y}_d}{a_2 \overline{y}_p + (b_2+ M_2) \overline{y}_d}, \,\,\,\,\,\,\,\,\,
\overline{d} = \frac{M_2 \overline{y}_d}{a_2 \overline{y}_p + (b_2+ M_2) \overline{y}_d},
\end{eqnarray}
\begin{eqnarray}\label{u12bar}
\overline{u}_p = \frac{a_1 \overline{y}_p}{a_1 \overline{y}_p + a_2 \overline{y}_d}, \,\,\,\,\,\,\,\,\,\,\,\, \overline{u}_d = \frac{a_2 \overline{y}_d}{a_1 \overline{y}_p + a_2 \overline{y}_d},
\end{eqnarray}
 \begin{eqnarray}\label{kappabar}
 \overline{k} = \frac{a_1 \overline{y}_p + a_2 \overline{y}_d}{\delta + \rho},
 \,\,\,\,\,\,\,\,\,\,\,\,
 \overline{c} = \frac{\sigma_1}{\sigma_2}(M_2-b_3)\overline{y}_d,
\end{eqnarray}
 \begin{eqnarray}\label{lambdabar}
 \overline{u} = \frac{1}{1-\sigma}\left(\overline{c}^{\sigma_1} \overline{c}^{\sigma_2} \right)^{1-\sigma},
 \,\,\,\,\,\,\,\,\,\,\,\,
 \overline{\lambda} = \frac{(1-\sigma)\sigma_1 \overline{u}}{\overline{c}},
\end{eqnarray}
where
\begin{eqnarray}\label{Pdelta}
P = 1-a_1 \frac{\delta + n}{\delta + \rho}, \,\,\,\,\,\,\,\,\,\, \Delta = P(M_2+b_2)-a_2 M_3.
\end{eqnarray}
\end{Proposition}

\begin{proof}
The proof is contained in Appendix \ref{propA}.
\end{proof}

\section{Variations of the production technology}

From now on, we will always assume that (\ref{assumptionsM}) holds, so that, according to Proposition \ref{propgrowthrate}, a unique balanced growth path exists. In the following, we shall investigate how changes in the technology levels impact the main economic variables in the steady state of this economy. More precisely, we are going to examine how a small variation of the parameters $A_p$ and $A_d$ affects the outputs $\overline{y}_p$, $\overline{y}_d$, $\overline{y}$ and the consumption $\overline{c}$.

\begin{Proposition}\label{propgrowthrateA}
We have: $\frac{\partial \overline{y}_p}{\partial A_p}>0$ and $\frac{\partial \overline{y}_d}{\partial A_p}<0$. Moreover, let us define:
\begin{eqnarray}\label{M4}
M_4 = P^2(M_2+b_2)-P(a_2M_3 + M_2M_3 + b_2M_3) + a_2M_3^2 +PM_2\frac{b_2+b_3}{b_3}(M_2+M_3+b_2).
\end{eqnarray}
\begin{eqnarray}\label{M5}
M_5 = M_2\frac{b_2+b_3}{b_3}\left(P(b_2+b_3)-a_2\left(1-\frac{\delta+n}{\delta+\rho}\right)\right)+(M_2-b_3)\Delta.
\end{eqnarray}
Then, $\frac{\partial \overline{y}}{\partial A_p}>0$ if $M_4>0$ and $\frac{\partial \overline{y}}{\partial A_p}<0$ if $M_4<0$. Finally, $\frac{\partial \overline{c}}{\partial A_p}>0$ if $M_5>0$ and $\frac{\partial \overline{c}}{\partial A_p}<0$ if $M_5<0$.
\end{Proposition}
\begin{proof}
The proof is contained in Appendix \ref{propB}.
\end{proof}

\begin{Proposition}\label{propgrowthrateB}
We have: $\frac{\partial \overline{y}_p}{\partial A_d}<0$ and $\frac{\partial \overline{y}_d}{\partial A_d}>0$. Moreover, let us define:
 \begin{eqnarray}\label{M6}
M_6 = P + a_2 - b_2 - b_3 - \left(1-a_2\frac{\delta+n}{\delta+\rho}\right),
 \,\,\,\,\,\,\,\,\,
 M_7 = a_2 \left(1-a_2\frac{\delta+n}{\delta+\rho}\right) -P(b_2+b_3).
\end{eqnarray}
Then, $\frac{\partial \overline{y}}{\partial A_d}>0$ if $M_6>0$ and $\frac{\partial \overline{y}}{\partial A_d}<0$ if $M_6<0$. Finally, $\frac{\partial \overline{c}}{\partial A_d}>0$ if $M_7>0$ and $\frac{\partial \overline{c}}{\partial A_d}<0$ if $M_7<0$.
\end{Proposition}

\begin{proof}
The proof is contained in Appendix \ref{propC}.
\end{proof}

Besides the expected effects of the total factor productivity parameters $A_d$ and $A_p$, we can identify the conditions for which a shock on the total factor productivity of either the physical sector (Proposition \ref{propgrowthrateA}) or the GE sector (Proposition \ref{propgrowthrateB}) causes an increase or a decrease in the total production and consumption levels. These results imply a possible change in time allocation and the level of digital leisure produced data. To analyse how work and leisure time choices change following shocks on the total factor productivity (TFP) (and other shocks) and how these impact the variable in the steady state equilibrium, we developed a series of simulations that we report in the following section.

\subsection{Simulations for changing TFPs}\label{simul_tfp}
To study the possible reallocation of time following a shock on total factor productivity, we use some numerical simulations. These simulations also allow us to study the impact on the steady-state equilibrium levels of production, consumption, and utility.\\
We first select a feasible and economically reasonable range of parameters, graphically investigating the monotonicity of the relationship between the variations of parameters and the level of each variable, finding that there is no change from positive to negative slope or vice versa within the range.\footnote{The graphs are available upon request.} This allows us to focus on a series of case studies for changing TFPs in the digital and the physical sector, which we report in Table \ref{tab:sim_tfp}. Notice that all simulations assume $\{\rho,n, \sigma_1,\sigma_2,\sigma,a_1,a_2,b_1,b_2,b_3,\delta\}=\{0.015,0.01,0.8,0.2,0.04,0.2,0.8,0.3,0.5,0.2,0.03\}$.\footnote{In these simulations, $M_4<0$.}

\begin{table}[ht]
\caption{Simulation for changing TFPs\label{tab:sim_tfp}}
\centering
\begin{tabular}{c|cc|cccccccccc}\hline\hline
Case & $A_p$ & $A_d$ & $\overline{h}_p$ & $\overline{h}_d$ & $\overline{y}_p$ & $\overline{y}_d$ & $\overline{y}$ & $\overline{c}$ & $\overline{d}$ & $\overline{u}$ & $\overline{u}_p$& $\overline{u}_d$ 
\\\hline
$S_{1,1}$ & 1 & 1 & 0.057 & 0.447 & 0.08 & 1.04 & 1.12 & 0.295 & 0.496 & 0.356 & 0.05 & 0.95 \\ 
$S_{1,2}$ & 1.02 & 1 & 0.093 & 0.410 & 0.138 & 0.977 & 1.12 & 0.317 & 0.497 & 0.377 & 0.09 & 0.91 \\ 
$S_{1,3}$ & 1 & 1.02 & 0.013 & 0.492 & 0.02 & 1.14 & 1.16 & 0.277 & 0.495 & 0.339 & 0.01 & 0.99 \\ 
\hline\hline
\end{tabular}
\end{table}

From Table \ref{tab:sim_tfp} we can observe how a shift in the physical sector (GE sector) total factor productivity parameter increases the working time by the same sector at the expense of the GE sector (physical sector), keeping the time spent for digital leisure almost stable. This is reflected in a shock on the two sectors' production, even though total production remains (approximately) stable. Interestingly, the intertemporal utility seems to be higher in case $S_{1,2}$, namely when the total factor productivity of the physical sector is relatively higher, possibly because both the labor and the capital allocated to this sector are higher (because of $\bar{u}_p$) than in the other two cases.

\subsection{Simulations for changing elasticity of output}
Following similar steps to those in \ref{simul_tfp}, we now investigate the impact of a change of $b_1$, $b_2$, and $b_3$, that is the elasticity of the output in the digital sector with respect to capital, labour, and leisure time, respectively, on the level of economic variables in the steady state equilibrium. In this case, since we do not have analytical results for the economic effects as in Propositions \ref{propgrowthrateA} and \ref{propgrowthrateB}, we also use the numerical simulations to show the impact on overall production and the production of the two sectors. Table \ref{tab:sim_b} reports the results of the simulation following different shocks on the elasticity of the digital sector's output with respect to its production factors. In all these simulations, we fixed $\{\rho,n,\sigma_1,\sigma_2,\sigma,a_1,a_2,A_p,A_d,\delta\} = \{0.015,0.01,0.5,0.5,0.4,0.6,0.4,1,1,0.01\}$.\footnote{In these simulations, $M_4>0$.}

\begin{table}[ht]
\caption{Simulation results for changing elasticities of digital output\label{tab:sim_b}}
\centering
\scalebox{0.7}{
\begin{tabular}{c|ccc|cccccccccc}\hline\hline
Case & $b_1$ & $b_2$ & $b_3$ & $\overline{h}_p$ & $\overline{h}_d$ & $\overline{y}_p$ & $\overline{y}_d$ & $\overline{y}$ & $\overline{c}$ & $\overline{d}$ & $\overline{u}$ & $\overline{u}_p$& $\overline{u}_d$ 
\\\hline
$S_{2,1}$ & 0.1 & 0.7 & 0.2 & 0.4347826 & 6.0e-9 & 51.119785 & 4.5e-7 & 51.1197856 & 13.2911443 & 0.56521739 & 3.05198304 & 0.9999999985 & 1.5e-9 \\ 
$S_{2,2}$ & 0.1 & 0.6 & 0.3 & 0.4347805 & 1.9e-6 & 51.119535 & 1.4e-4 & 51.1196812 & 13.2911287 & 0.56521766 & 3.05198241 & 0.9999995238 & 4.7e-7 \\ 
$S_{2,3}$ & 0.1 & 0.5 & 0.4 & 0.4347427 & 2.9e-5 & 51.115099 & 2.7e-3 & 51.1178283 & 13.2908536 & 0.56522823 & 3.05198059 & 0.9999911023 & 8.9e-6 \\ 
\hline
$S_{2,4}$ & 0.3 & 0.5 & 0.2 & 0.4347791 & 2.9e-6 & 51.119375 & 2.7e-4 & 51.1196490 & 13.2911307 & 0.56521797 & 3.05198059 & 0.9999973217 & 2.7e-6 \\ 
$S_{2,5}$ & 0.3 & 0.4 & 0.3 & 0.4346324 & 1.0e-4 & 51.102121 & 1.2e-2 & 51.1138977 & 13.2905556 & 0.56526747 & 3.05202360 & 0.9998847890 & 1.2e-4 \\ 
$S_{2,6}$ & 0.3 & 0.3 & 0.4 & 0.4330601 & 5.2e-4 & 50.917258 & 8.2e-2 & 51.0628249 & 13.2879798 & 0.56601137 & 3.05305048 & 0.9980976972 & 8.1e-4 \\ 
\hline
$S_{2,7}$ & 0.5 & 0.3 & 0.2 & 0.4291400 & 3.3e-3 & 50.456352 & 0.52 & 50.9736055 & 13.2945177 & 0.56756051 & 3.05600583 & 0.9915294368 & 8.4e-3 \\ 
$S_{2,8}$ & 0.5 & 0.2 & 0.3 & 0.4103029 & 9.5e-3 & 48.241572 & 2.24 & 50.4856033 & 13.3057793 & 0.58015416 & 3.07697429 & 0.9626827758 & 3.7e-2 \\ 
$S_{2,9}$ & 0.5 & 0.1 & 0.4 & 0.3739289 & 1.2e-2 & 43.964875 & 5.58 & 49.5432802 & 13.3275252 & 0.61420982 & 3.13161691 & 0.9043749912 & 9.6e-2 \\ 
\hline\hline
\end{tabular}
}
\end{table}

From Table \ref{tab:sim_b}, we can see how a technological change that modifies the elasticity of the product with respect to data (namely, $b_3$) clearly has an impact on time allocation: the consumer increases his leisure time and the time he works in the digital sector as $b_3$ increases, and this effect is stronger for $d$ and weaker for $h_d$ as $b_1$ increases. Digital leisure and GE sector working time crowd out physical sector working time as the importance of data in the production function of the GE sector increases.

\section{Conclusions}\label{conc}
In this paper, we investigated how the digitization of the market and the emergence of the GE influence economic growth in the long run by setting a two-sector growth model with time allocation choice. Considering the impact on labor and leisure and the consequent change in the way people make their choice of consumption, we took into account the role of data as a by-product of the leisure activity of the consumers. We observed how a shift in the physical (GE) sector total factor productivity parameter increases the working time allocated to this sector by the workers and decreases the working time spent in the GE (physical) sector, keeping the time spent for digital leisure almost stable. Even though total production remains stable, when the total factor productivity of the physical sector is relatively higher, the intertemporal utility seems to be higher in some cases. On the other hand, we saw how a technological change in product elasticity with respect to data increases both the leisure time of the consumer and its working time in the digital sector increase. The impact of these shocks on time allocation allowed us to find a novel result for this economy: the available time allocated to digital leisure and the work in the GE sector crowds out the working time spent in the physical sector as the importance of data in the production function of the GE sector increases. \\
Our findings contribute to highlighting the relevance of data in the current market setting, where digital firms are becoming more and more important and flank traditional firms, while better exploiting the big data availability due to the increased use of devices and digital platforms. We demonstrate that the importance that data has as a production factor can modify not only the firm-side variables but also the choices the workers make. Understanding the role of data, an increasingly decisive factor in the digital and non-digital sectors in the last years, is also a necessity to design possible policies to help workers and firms adapt to the transition to this new type of economy where mechanisms of substitution between free time and working time allocated to the digital and non-digital sectors are at work.\\
Our model could be developed along several lines, considering for example different types of human capital needed to work in either one sector or the other (or both), which could increase the impact of data on production in the digital sector or restrain it, depending on whether the workers are well-trained or skilled to operate in such a sector. Another interesting extension of our analysis is to enrich the model considering market prices and study how they can be impacted by shocks of productivity and production elasticities. \\
The framework proposed in this paper could also guide the decisions of agents operating in a market with both non-digital and digital firms, e.g., setting a limit to data usability along the lines of the General Data Protection Regulation (GDPR) or a constraint to the additional GE jobs persons are allowed to make (e.g., an exclusivity contract between a traditional sector firm and a worker, which then is limited in spending her/his time in the GE sector).\\
Our model also provides implications for managers of traditional and digital firms as investment strategies in digital technologies require both soft- and hard-skilled workers. While the former can be hired from the job market with unregulated contracts, the latter appear to be workers enrolled in collective labor agreements and protected by trade unions with higher wages. Moreover, the possibility of collecting consumer data without paying for them allows managers to invest resources in more efficient and expensive digital technologies, thus accelerating the digital transition process. In spite of that, competition among digital firms to acquire the most valuable consumer data by providing compensation for them could increase the cost of data collection and slow down the process.

\bibliographystyle{unsrtnat}
\bibliography{references}

\appendix

\section{Proofs of Proposition \ref{propgrowthrate}}\label{propA}
\setcounter{equation}{0}
\renewcommand{\theequation}{\Alph{section}.\arabic{equation}}
From equation (\ref{Hup}) we obtain
\begin{eqnarray}\label{Hupbis}
\frac{u_p}{a_1 y_p}- \frac{u_d}{b_1 y_d}=0,
\end{eqnarray}
which, together with the fact that $u_p+u_d=1$, yields
\begin{eqnarray}\label{u12}
u_p = \frac{a_1 y_p}{a_1 y_p + a_2 y_d}, \,\,\,\,\, u_d = \frac{a_2 y_d}{a_1 y_p + a_2 y_d}.
\end{eqnarray}
By equating the left-hand side of (\ref{Hhp}) with the left-hand side of (\ref{Hhd}) we obtain
\begin{eqnarray}\label{hd}
h_d = \frac{b_2 y_d h_p}{a_2 y_p},
\end{eqnarray}
and, by substituting (\ref{hd}) in (\ref{Hupbis}) we compute
\begin{eqnarray}\label{hp0}
h_p = \frac{a_2 y_p}{a_2 y_p + (b_2+b_3) y_d + (1-\sigma)\sigma_2 \frac{u}{\lambda}},
\end{eqnarray}
which, taking into account (\ref{Hc}) and the fact that $\sigma_1 + \sigma_2=1$, is rewritten as follows:
\begin{eqnarray}\label{hp1}
h_p = \frac{a_2 y_p}{a_2 y_p + (b_2+b_3) y_d + \frac{\sigma_2}{\sigma_1}c}.
\end{eqnarray}
Furthermore, (\ref{hd}) and (\ref{hp1}) yield
\begin{eqnarray}\label{hd1}
h_d = \frac{b_2 y_d}{a_2 y_p + (b_2+b_3) y_d + \frac{\sigma_2}{\sigma_1}c}.
\end{eqnarray}
Then, based on (\ref{hp1}) and (\ref{hd1}), we can compute
\begin{eqnarray}\label{hdhp}
1 - h_p - h_d = \frac{b_3 y_d + \frac{\sigma_2}{\sigma_1}c}{a_2 y_p + (b_2+b_3) y_d + \frac{\sigma_2}{\sigma_1}c}.
\end{eqnarray}
Moreover, at the steady state, $\dot{\lambda} = 0$, which yields
 \begin{eqnarray}\label{kappa}
k = \frac{a_1 y_p + a_2 y_d}{\delta + \rho}.
\end{eqnarray}
By substituting (\ref{hp1}), (\ref{hd1}), (\ref{hdhp}) and (\ref{kappa}) in (\ref{ypsmall}), and by using the first of relations (\ref{cobbab}), we obtain
\begin{eqnarray}\label{y1}
\left({a_2 y_p + (b_2+b_3) y_d + \frac{\sigma_2}{\sigma_1}c}\right)^{a_2}=\frac{A_p a_1^{a_1} a_2^{a_2}}{(\delta+\rho)^{a_1}}.
\end{eqnarray}
Similarly, by substituting (\ref{hp1}), (\ref{hd1}), (\ref{hdhp}) and (\ref{kappa}) in (\ref{ydsmall}), and by using the second of relations (\ref{cobbab}), we obtain
\begin{eqnarray}\label{y2}
\left({a_2 y_p + (b_2+b_3) y_d + \frac{\sigma_2}{\sigma_1}c}\right)^{(b_2+b_3)}=\frac{A_d b_1^{b_1} b_2^{b_2}\left(b_3 y_d + \frac{\sigma_2}{\sigma_1}c\right)^{b_3}}{(\delta+\rho)^{b_1} y_d^{b_3}}.
\end{eqnarray}
Relation (\ref{y1}) can be rewritten as follows:
\begin{eqnarray}\label{y1bis}
{a_2 y_p + (b_2+b_3) y_d + \frac{\sigma_2}{\sigma_1}c} = M_1,
\end{eqnarray}
where $M_1$ is defined in (\ref{M1}). Substitution of (\ref{y1bis}) in (\ref{y2}) yields
\begin{eqnarray}\label{y2bis}
\left(\frac{b_3 y_d + \frac{\sigma_2}{\sigma_1}c}{y_d}\right)^{b_3} = M_1^{b_2+b_3} \frac{(\delta+\rho)^{b_1}}{A_d b_1^{b_1} b_2^{b_2}},
\end{eqnarray}
or, equivalently,
\begin{eqnarray}\label{y2tris}
b_3 y_d + \frac{\sigma_2}{\sigma_1}c = M_2 y_d,
\end{eqnarray}
where $M_2$ is defined in (\ref{M1}). From (\ref{y2tris}), let us compute
\begin{eqnarray}\label{cM2}
c = \frac{\sigma_1}{\sigma_2}(M_2-b_3) y_d,
\end{eqnarray}
and let us substitute (\ref{cM2}) in (\ref{y1bis}), obtaining
\begin{eqnarray}\label{y1tris}
a_2 y_p + (b_2+ M_2) y_d = M_1.
\end{eqnarray}
Since, at the steady state, we have $\dot{k}=0$, (\ref{ypsmall}), (\ref{ydsmall}) and (\ref{dynamicsk}) yield
\begin{eqnarray}\label{y2quadris}
y_p + y_d - c - (\delta + n)k = 0,
\end{eqnarray}
which, by using the third of (\ref{M1}), (\ref{kappa}), (\ref{cM2}) and the first of (\ref{Pdelta}), is rewritten as follows:
\begin{eqnarray}\label{y2quinque}
P y_p +M_3 y_d = 0.
\end{eqnarray}
Relations (\ref{y2quinque}) and (\ref{y1tris}) form a system of two linear equations in the unknowns $y_p$ and $y_d$, whose determinant is equal to $\Delta$, as defined in the second of (\ref{Pdelta}). Relation (\ref{assumptionsM}) and $\rho>n$, and the fact that $0<a_1<1$ imply $\Delta>0$, and thus we have a unique solution $y_p = \overline{y}_p$ and $y_d = \overline{y}_d$, where $\overline{y}_p$ and $\overline{y}_d$ are given by (\ref{ypbarydbar}).\\
To this point, the third of (\ref{ypbarydbar}) follows from (\ref{ytotalsmall}), (\ref{kappabar}) follows from (\ref{kappa}) and (\ref{cM2}), and (\ref{u12bar}) follows from (\ref{u12}). Then, (\ref{hpbar}) follows from (\ref{hp1}), (\ref{hd1}), (\ref{hdhp}) and (\ref{cM2}). Finally, the first of (\ref{lambdabar}) follows from (\ref{uagent}), whereas the second of (\ref{lambdabar}) is obtained by setting $\dot{\lambda} = 0$ in (\ref{Hlambda}). \\
The fact that assumptions (\ref{assumptionsM}) must be satisfied in order to have at least one economically meaningful steady state solution is very simple to prove. First of all, if the first of (\ref{assumptionsM}) does not hold, then, according to (\ref{cM2}), consumptions could not be strictly positive (notice that the case $\overline{c} = 0$ must be excluded, since, upon (\ref{uagent}), it would yield $\overline{u} = 0$, which does not amount to maximizing (\ref{totalutility})). Moreover, the second of (\ref{assumptionsM}) must hold true as well. In fact, according to (\ref{y2quinque}) and to the fact that $a_1 < 1$ and $\rho > n$, we can have $\overline{y}_p \geq 0$ and $\overline{y}_d \geq 0$ only if the second of (\ref{assumptionsM}) is satisfied.\\
Finally, it is clear that $\overline{y}_p$ and $\overline{y}_d$ must satisfy the system of equations (\ref{y1tris}) and (\ref{y2quinque}). Therefore, if assumptions (\ref{assumptionsM}) hold true, $\overline{y}_p$ and $\overline{y}_d$ exist unique, and thus, according to the above calculations, there exists a unique steady state solution of (\ref{Hc})-(\ref{Hlambda}) satisfying (\ref{constrainth}), the constraints on $u_p$ and $u_d$, (\ref{ypsmall}), (\ref{ydsmall}), (\ref{ytotalsmall}), (\ref{dynamicsk}) and (\ref{additionalrequirements}).

\section{Proofs of Proposition \ref{propgrowthrateA}}\label{propB}
By taking derivatives of $y_p$ as in (\ref{ypbarydbar}) with respect to $A_p$, we obtain:
\begin{eqnarray}\label{dypdA}
\frac{\partial \overline{y}_p}{\partial A_p} = \frac{M_1}{a_2 A_p \Delta^2} \left( -M_3(PM_2 + Pb_2-a_2M_3)
+ PM_2(M_2+M_3+b_2) \frac{b_2+b_3}{b_3}\right).
\end{eqnarray}
From the third of $(\ref{M1})$ it immediately follows that $M_2 + M_3 >0$, and thus, since $M_3<0$, it must be
 $\frac{\partial \overline{y}_p}{\partial A_p}>0$.
By taking derivatives of $y_d$ as in (\ref{ypbarydbar}) with respect to $A_p$, we obtain:
\begin{eqnarray}\label{dyddA}
\frac{\partial \overline{y}_d}{\partial A_p} = \frac{P M_1}{a_2 A_p \Delta^2} \left(PM_2+Pb_2-a_2M_3-M_2(P+a_2) \frac{b_2+b_3}{b_3}\right),
\end{eqnarray}
which, by straightforward algebra, can be rewritten as follows:
\begin{eqnarray}\label{dyddAbis}
\frac{\partial \overline{y}_d}{\partial A_p} = \frac{P M_1}{a_2 A_p \Delta^2} \left(Pb_2-a_2M_3-\frac{b_2}{b_3}M_2(P+a_2)-M_2a_2\right),
\end{eqnarray}
and, using the third of (\ref{M1}), simplifies to
\begin{eqnarray}\label{dyddAtris}
\frac{\partial \overline{y}_d}{\partial A_p} = \frac{P M_1}{a_2 A_p \Delta^2} \left(Pb_2-\frac{b_2}{b_3}M_2(P+a_2)-a_2\left(a-a_2\frac{\delta+n}{\delta+\rho}+b_3\right)\right).
\end{eqnarray}
Then, since $M_2 > b_3$ (according to the first of (\ref{assumptionsM})), we immediately conclude that $\frac{\partial \overline{y}_d}{\partial A_p}<0$.\\
Summing up (\ref{dypdA}) and (\ref{dyddA}) and using (\ref{M4}) we can easily obtain:
\begin{eqnarray}\label{dysumdAtris}
\frac{\partial \overline{y}}{\partial A_p} = \frac{P M_1 M_4}{a_2 A \Delta^2},
\end{eqnarray}
and thus it follows that $\frac{\partial \overline{y}}{\partial A_p}>0$ if $M_4>0$ and $\frac{\partial \overline{y}}{\partial A_p}<0$ if $M_4<0$.\\
Finally, let us analyze consumption. By differentiating (\ref{cM2}) with respect to $A_p$ we can compute
 \begin{eqnarray}
\frac{\partial \overline{c}}{\partial A_p} = \frac{b_2+b_3}{b_3} \frac{M_2}{a_2 A_p} y_d + (M_2-b_3) \frac{\partial \overline{y}_d}{\partial A_p},
 \end{eqnarray}
which, by using the second of (\ref{ypbarydbar}) and (\ref{dyddA}) (and some simple algebra), is rewritten as follows:
 \begin{eqnarray}
\frac{\partial \overline{c}}{\partial A_p} = \frac{P M_1}{a_2 A_p \Delta^2} \left(M_2 \frac{b_2+b_3}{b_3} \left(
\Delta-(M_2-b_3)(P+a_2)\right) + (M_2-b_3)\Delta \right).
 \end{eqnarray}
Then, by substituting the expression of $\Delta$ in the second of (\ref{Pdelta}) and by applying (\ref{M5}), we obtain
 \begin{eqnarray}
\frac{\partial \overline{c}}{\partial A_p} = \frac{P M_1 M_5}{a_2 A_p \Delta^2},
 \end{eqnarray}
and thus it follows that $\frac{\partial \overline{c}}{\partial A_p}>0$ if $M_5>0$ and $\frac{\partial \overline{c}}{\partial A_p}<0$ if $M_5<0$.

\section{Proofs of Proposition \ref{propgrowthrateB}}\label{propC}
By taking derivatives of $y_p$ as in (\ref{ypbarydbar}) with respect to $A_d$, we obtain:
\begin{eqnarray}\label{dypdB}
\frac{\partial \overline{y}_p}{\partial A_d} = -\frac{P M_1 M_2}{b_3 A_d \Delta^2} \left(M_2 + M_3 + b_2\right).
\end{eqnarray}
From the third of $(\ref{M1})$ it immediately follows that $M_2 + M_3 >0$, and thus it must be $\frac{\partial \overline{y}_p}{\partial A_d}<0$.
By taking derivatives of $y_d$ as in (\ref{ypbarydbar}) with respect to $A_d$, we obtain:
\begin{eqnarray}\label{dyddB}
\frac{\partial \overline{y}_d}{\partial A_d} = \frac{P M_1 M_2}{b_3 A_d \Delta^2}(P+a_2),
\end{eqnarray}
which implies that $\frac{\partial \overline{y}_d}{\partial A_d}>0$.
Summing up (\ref{dypdB}) and (\ref{dyddB}) and using the third of (\ref{M1}) and the first of (\ref{M6}), we obtain:
\begin{eqnarray}\label{dysumdBtris}
\frac{\partial \overline{y}}{\partial A_d} = \frac{P M_1 M_2}{b_3 A_d \Delta^2} M_6,
\end{eqnarray}
and thus it follows that $\frac{\partial \overline{y}}{\partial A_d}>0$ if $M_6>0$ and $\frac{\partial \overline{y}}{\partial A_d}<0$ if $M_6<0$.
Finally, by differentiating (\ref{cM2}) with respect to $A_d$, we compute
 \begin{eqnarray}
\frac{\partial \overline{c}}{\partial A_d} = \frac{P M_1 M_2}{b_3 A_d \Delta^2} \left((M_2-b_3)(P+a_2)-P(M_2+b_2) + a_2M_3 \right),
 \end{eqnarray}
 which, by using the third of (\ref{M1}), the second of (\ref{M6}) and some simple algebra is simplified as follows:
 \begin{eqnarray}
\frac{\partial \overline{c}}{\partial A_d} = \frac{P M_1 M_2 M_7}{b_3 A_d \Delta^2}.
 \end{eqnarray}
It then follows that $\frac{\partial \overline{c}}{\partial A_d}>0$ if $M_7>0$ and $\frac{\partial \overline{c}}{\partial A_d}<0$ if $M_7<0$.

\end{document}